\documentclass[sn-nature]{sn-jnl}% Style for submissions to Nature Portfolio journals
\usepackage{graphicx}%
\usepackage{multirow}%
\usepackage{amsmath,amssymb,amsfonts}%
\usepackage{amsthm}%
\usepackage{mathrsfs}%
\usepackage[title]{appendix}%
\usepackage{xcolor}%
\usepackage{textcomp}%
\usepackage{manyfoot}%
\usepackage{booktabs}%
\usepackage{algorithm}%
\usepackage{algorithmicx}%
\usepackage{algpseudocode}%
\usepackage{listings}%
\usepackage{romannum}
\usepackage{pdflscape}

\usepackage{amsmath,siunitx}

\usepackage{lmodern}
\usepackage{anyfontsize}
\usepackage{comment}

\begin{document}

\title[Super-Chandrasekhar mass DWD at 49\,pc]{A super-Chandrasekhar mass type Ia supernova progenitor at 49\,pc set to detonate in 23\,Gyr}


\author*[1]{\fnm{James} \sur{Munday}}\email{james.munday98@gmail.com}
\author[2]{\fnm{Ruediger} \sur{Pakmor}}
\author[1]{\fnm{Ingrid} \sur{Pelisoli}}
\author[3,4,5]{\fnm{David} \sur{Jones}}
\author[1]{\fnm{Snehalata} \sur{Sahu}}
\author[1]{\fnm{Pier-Emmanuel} \sur{Tremblay}}
\author[2]{\fnm{Abinaya Swaruba} \sur{Rajamuthukumar}}
\author[6,7,8]{\fnm{Gijs} \sur{Nelemans}}
\author[1]{\fnm{Mark} \sur{Magee}}
\author[9]{\fnm{Silvia} \sur{Toonen}}
\author[1]{\fnm{Antoine} \sur{B\'edard}}
\author[10]{\fnm{Tim} \sur{Cunningham}}

\affil*[1]{\orgdiv{Department of Physics}, \orgname{University of Warwick}, \orgaddress{\street{Gibbet Hill Road}, \city{Coventry}, \postcode{CV4 7AL}, \country{United Kingdom}}}
\affil[2]{\orgdiv{Max-Planck-Institut für Astrophysik}, \orgname{Max-Planck-Institut für Astrophysik}, \orgaddress{\street{Karl-Schwarzschild-Str. 1}, \postcode{D-85748}, \country{Germany}}}
\affil[3]{\orgdiv{Instituto de Astrof\'isica de Canarias}, \orgaddress{\city{La Laguna}, \postcode{E-38205}, \state{Tenerife}, \country{Spain}}}
\affil[4]{\orgdiv{Departamento de Astrof\'isica}, \orgname{Universidad de La Laguna}, \orgaddress{\street{E-38206}, \city{La Laguna}, \postcode{610101}, \state{Tenerife}, \country{Spain}}}
\affil[5]{\orgdiv{Nordic Optical Telescope}, \orgname{Rambla Jos\'e Ana Fern\'andez P\'erez}}
\affil[6]{\orgdiv{Department of Astrophysics/IMAPP}, \orgname{Radboud University}, \orgaddress{\street{P.O. Box 9010}, \postcode{6500 GL Nijmegen}, \country{The Netherlands}}}
\affil[7]{\orgdiv{Institute for Astronomy}, \orgname{KU Leuven}, \orgaddress{\street{Celestijnenlaan 200D}, \city{Leuven}, \country{Belgium}}}
\affil[8]{\orgname{SRON, Netherlands Institute for Space Research}, \orgaddress{\street{Niels Bohrweg 4,
2333 CA Leiden}, \country{The Netherlands}}}
\affil[9]{\orgdiv{Anton Pannekoek Institute for Astronomy}, \orgname{University of Amsterdam}, \orgaddress{\street{1090 GE Amsterdam}, \country{The Netherlands}}}
\affil[10]{\orgdiv{Center for Astrophysics}, \orgname{Harvard \& Smithsonian}, \orgaddress{\street{60 Garden Street}, \city{Cambridge}, \postcode{MA 02138}, \country{USA}}}


\abstract{Double white dwarf binaries are a leading explanation to the origin of type \Romannum{1}a supernovae, but no system exceeding the Chandrasekhar mass limit (1.4\,M$_\odot$) has been found that will explode anywhere close to a Hubble time. Here, we present the super-Chandrasekhar mass double white dwarf WDJ181058.67+311940.94 whose merger time ($22.6\pm1.0$\,Gyr) is of the same order as a Hubble time. The mass of the binary is large, combining to $1.555\pm0.044$\,M$_\odot$,  while being located only 49\,pc away. We predict that the binary will explode dynamically via a double detonation destroying both stars just before they merge, appearing as a subluminous type Ia supernova with a peak apparent magnitude of about $m_V=-16$ (200\,000 times brighter than Jupiter). The observationally-derived birthrate of super-Chandrasekhar mass double white dwarfs is now at least $6.0\times10^{-4}$\,yr$^{-1}$ and the observed rate of type \Romannum{1}a supernovae in the Milky Way from such systems is approximately $4.4\times10^{-5}$\,yr$^{-1}$, while the predicted type \Romannum{1}a supernova rate in the Milky Way from all progenitor channels is about sixty times larger. Hence, WDJ181058.67+311940.94 mitigates the observed deficit of massive double white dwarfs witnessed in volume-complete populations, but further evidence is required to determine the majority progenitors of type \Romannum{1}a supernovae.}

\maketitle

\section{Introduction}\label{sec1}
Binaries comprising at least one white dwarf are the progenitors of type \Romannum{1}a supernovae \citep{Nugent2011, Bloom2012}. Type \Romannum{1}a supernovae show an absence of hydrogen in their spectrum and are caused by the thermonuclear explosion of a carbon-oxygen white dwarf. Nuclear fusion transforms a significant amount of, or the entire white dwarf, into heavier elements and ejects them into the interstellar medium. However, the stellar type of the companion to the white dwarf in type \Romannum{1}a progenitors remains largely unclear \citep[e.g.][]{Maoz2012supernovaReview, Liu2023supernovaReview, Soker2024}.

The substantial population size of double white dwarf binaries has naturally led to them being one of the leading progenitor candidates to explain the abundance of type Ia supernovae \citep{Webbink1984DWDprogenitorsRCrB, IbenTutukov1984}. These systems form on compact orbits with an orbital period on the timescale of hours to days \citep{Nelemans2001closeWDs} (orbital separations of hundredths to tenths of astronomical units) following a series of mass transfer events \citep{PostnovAndYungelson2014}. The gradual loss of orbital angular momentum through gravitational wave radiation draws the two stars closer until the orbital period of massive double white dwarfs is a couple of minutes, initiating unstable mass transfer and leading to the demise of the system \citep{Ruiter2011}.

Although many compact double white dwarfs have been discovered on the brink of coalescence \citep[e.g.][]{Burdge2020systematic, Brown2020elmNorthFinal, Ren2023}, we have had no direct evidence that these systems exist in nearby, volume-complete populations \citep{Toonen2017, Hollands2018, OBien2024}, casting doubt on whether double white dwarfs can account for a large percentage of the observed type \Romannum{1}a supernova rates. Current synthetic models of the population indicate that super-Chandrasekhar mass limit double white dwarfs are indeed suspected to be scarce \citep{Toonen2012type1aCommonEnvelope, RebassaMansergas2019, Li2023}. However, based on the models of \cite{Toonen2012type1aCommonEnvelope} we expect about 150 compact double white dwarf binaries to have total masses that exceed 1.5\,M$_\odot$ within 100\,pc, about one quarter of which merge in under a Hubble time. There has been only one super-Chandrasekhar mass double white dwarf binary discovered \citep[NLTT 12758,][]{Kawka2017}, but its 1.15\,d period means that the two stars will come into contact in about 10 Hubble times. There are a handful of other candidate subluminous type \Romannum{1}a progenitors that are double white dwarfs that have total masses smaller than the Chandrasekhar mass limit \citep[e.g.][]{Maxted2002, 2003ANapiwotzki2003, Nelemans2005, RebassaMansergas2017, Munday2023, Munday2024dbl}, two white dwarf+hot subdwarf systems that exceed 1.4\,$\mathrm{M}_\odot$ and have an impending supernova fate \citep{Pelisoli2021type1aWDsubdwarf, Luo2024superChandWDsubdwarf} and one other white dwarf+hot subdwarf that is also a strong candidate \citep{Maxted2000kpd, Geier2007}.

Growing observational evidence supports hot subdwarfs as some of the products of binary evolution \citep{Pelisoli2020}, but, although more super-Chandrasekhar mass systems have been discovered, the binaries much less densely populate the Galaxy \citep{Dawson2024}. The observed rate of type \Romannum{1}a supernovae initiated from the white dwarf+hot subdwarf channel is expected to be at least $(1.5$--$7)\times10^{-5}$\,yr$^{-1}$ \citep{Pelisoli2021type1aWDsubdwarf}, while the rate of type Ia supernovae in the Galaxy from all progenitors is about $2.8\pm0.6\times10^{-3}$\,yr$^{-1}$ \citep{LiSupernovaeRates2011,Maoz2014Type1aProgenitors, Li2023, Liu2023supernovaReview} as inferred through observations of explosions in other galaxies of similar redshift. Multiple other evolutionary scenarios have been suggested as causes for normal and peculiar type \Romannum{1}a supernovae \citep[][]{Liu2023supernovaReview} having different companion compositions, but the extent to which they contribute towards the missing fraction of type \Romannum{1}a supernovae is unclear. This ambiguity on the nature of type \Romannum{1}a progenitors is cosmologically problematic. A primary reason is that, until we confirm the leading progenitors of a type \Romannum{1}a, systematic errors to the distances derived to other galaxies could lead to inaccurate measurements, which is particularly troublesome for galaxies at high redshifts \citep{Pan2012, Maoz2014Type1aProgenitors}. In addition, the details of the ejecta velocity and its constituents are important for star formation \citep{Lacchin2021} and the dynamics of gas in galaxies \citep{Jimenez2015}. Not only does the discovery of a local, compact, super-Chandrasekhar mass double white dwarf have the ability to resolve the dearth of systems in the observed sample, but a sample of such systems has the power to reduce the uncertainty of this cosmologically fundamental event.

\begin{landscape}
\begin{figure*}
    \centering
    \includegraphics[keepaspectratio, trim={0.35cm 0cm 1.25cm 1.8cm},clip,width=6.4cm]{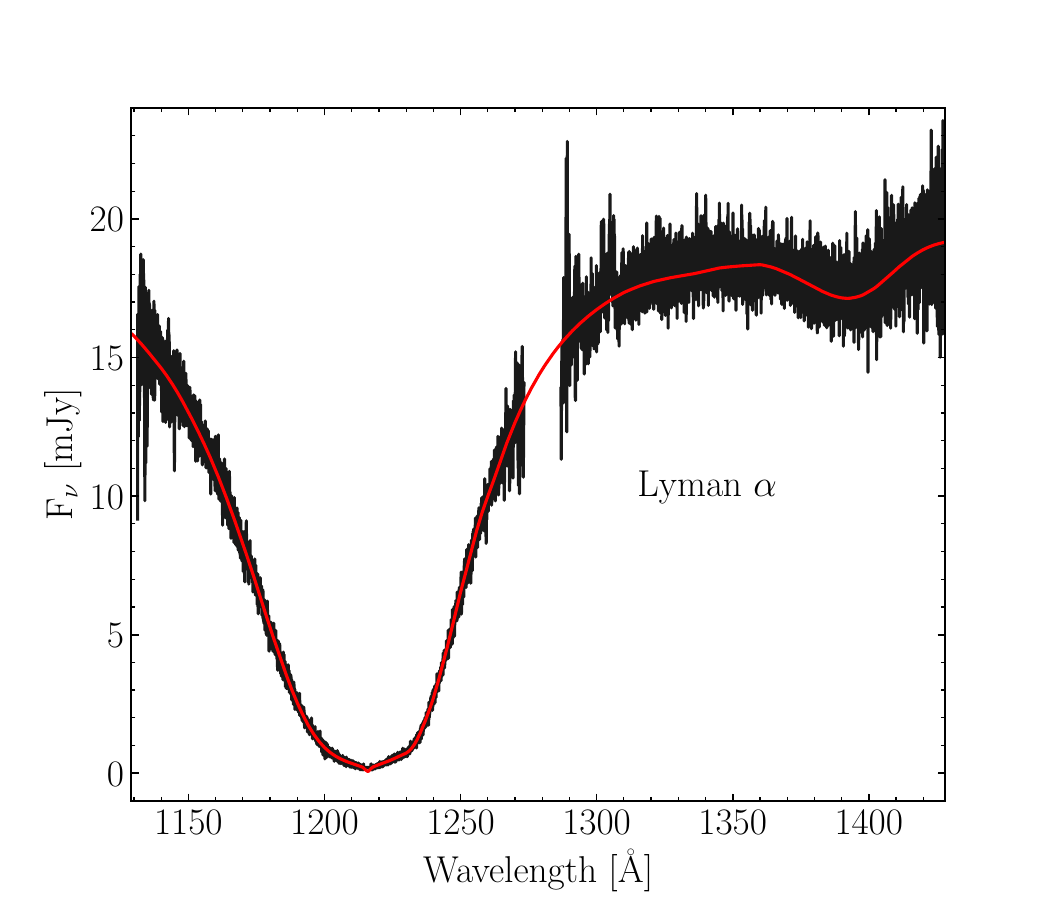}
    \includegraphics[keepaspectratio, trim={0.35cm 0cm 1.25cm 1.8cm},clip,width=6.4cm]{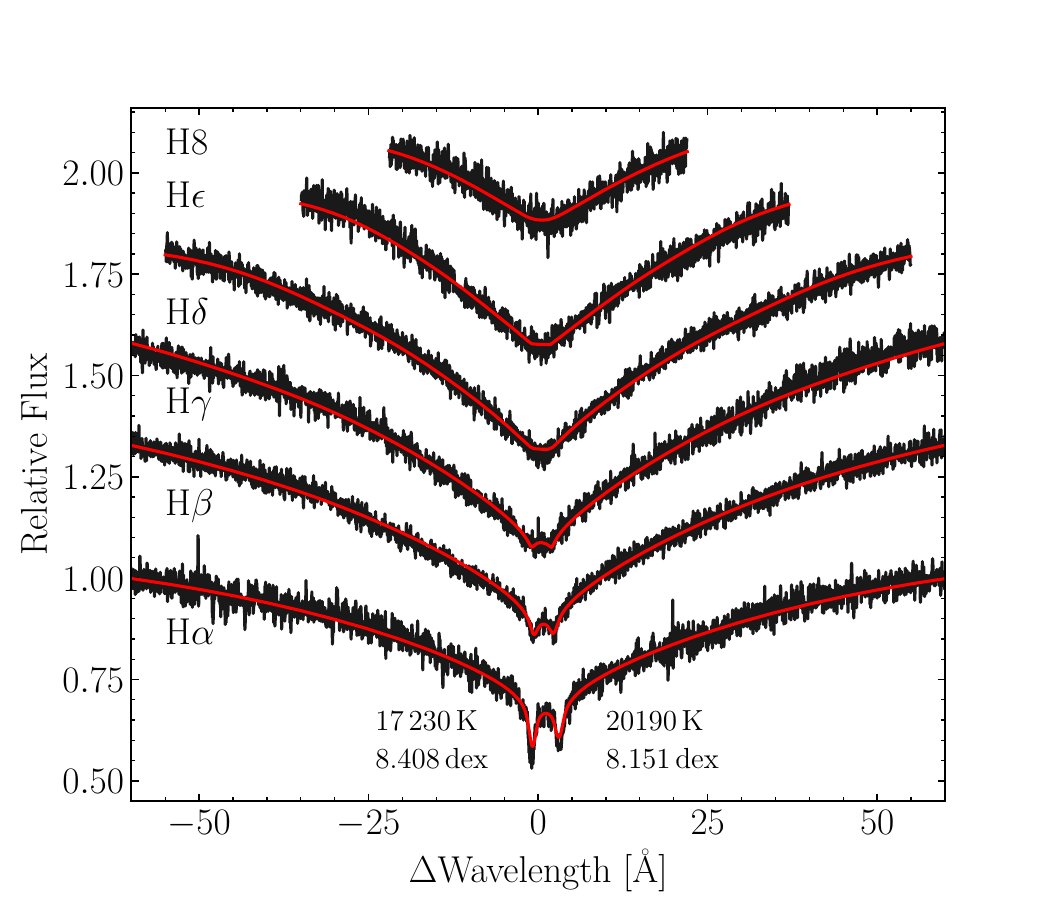}
    \includegraphics[keepaspectratio, trim={0.5cm 0cm 1.1cm 1.25cm},clip,width=6.4cm]{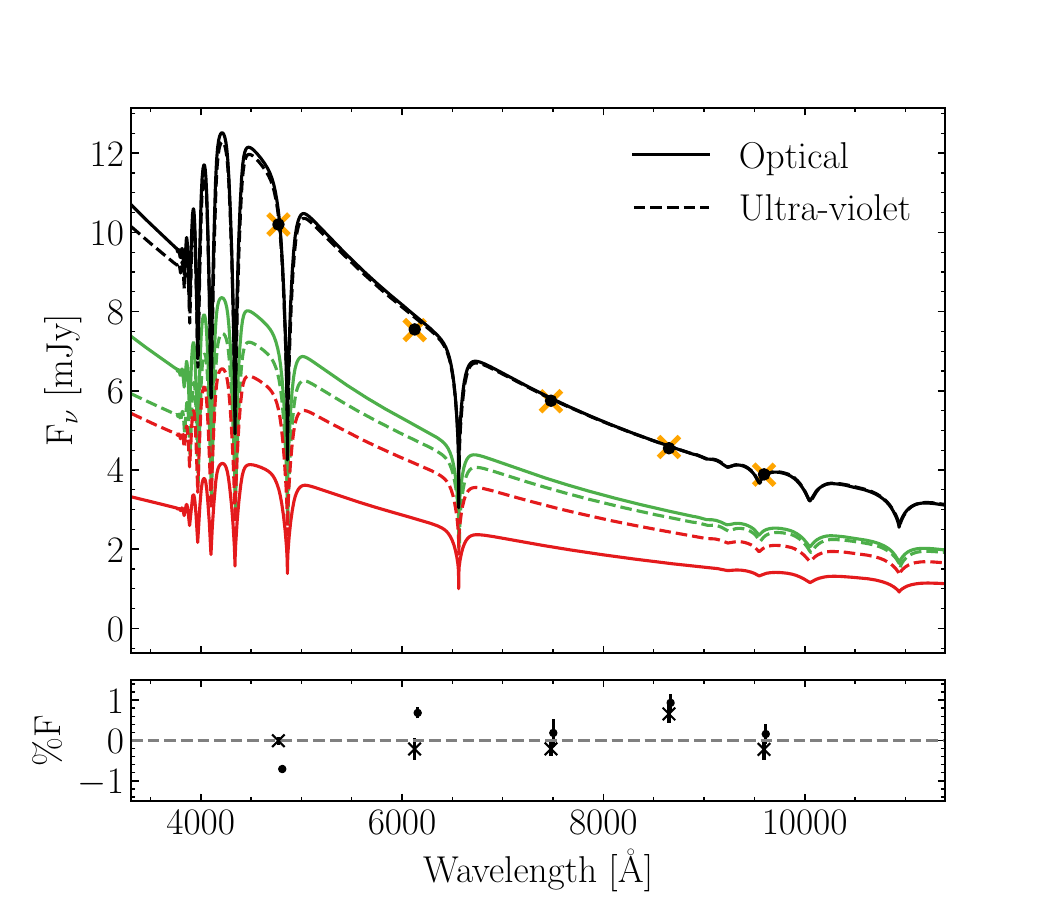}
    \caption{\textbf{Atmospheric fits to the photometric and spectroscopic datasets.} \textit{Left:} The HST/COS ultra-violet spectrum with the synthetic spectrum from the hybrid HST/COS with Pan-STARRS photometry fit for a two-star model overlaid in red. The corresponding atmospheric parameters of the DA white dwarfs are $T_{\textrm{eff},1}=18\,630$\,K, $\log(\textrm{g}_1)=8.307$\,dex, $T_{\textrm{eff},2}=18\,010$\,K, $\log(\textrm{g}_2)=8.178$\,dex. \textit{Middle:} A single UVES spectrum from H$\alpha$ to H8 with the synthetic spectral model for atmospheric parameters $T_{\textrm{eff},1}=17\,230$\,K, $\log(\textrm{g}_1)=8.408$\,dex, $T_{\textrm{eff},2}=20\,190$\,K, $\log(\textrm{g}_2)=8.151$\,dex, overplotted in red. We remind the reader that all Balmer lines up to H11 were fit but are omitted from the plot for clarity. \textit{Right:} The observed fluxes in Pan-STARRS (black circles) and the synthetic photometry in each filter for the same atmospheric parameters (orange crosses). The percentage flux residual between the data and the combined flux is found below. The flux contributed from the more massive (red) and less massive (green) stars are included for the ultra-violet (dashed) and optical (solid line) fits. The \textit{Gaia} parallax measurement with a Gaussian prior was an independent variable to scale the observations from an Eddington to an absolute flux perceived in the solar system}
    \label{fig:SpecPhotFit}
\end{figure*}
\end{landscape}

\begin{table*}
    \centering
    \caption{\textbf{The atmospheric parameters for each spectroscopic data set.} Hybrid fitting was performed in all cases using Pan-STARRS photometry. A systematic difference between the ultra-violet spectroscopy and the optical photometry was considered in the fitting (Section~\ref{subsec:AtmosphericUV}). Masses are inferred by interpolation of evolutionary sequences \citep{Bedard2020} and $M_\textrm{T}$ is the total mass of the system. The final adopted values were obtained by concatenating the distributions obtained for each parameter to then quote the median and 68\% confidence interval on the T$_\textrm{eff}$ and $\log(g)$, while interpolating to find masses. The more/less massive star is labelled with subscript 1/2 respectively. The WHT/ISIS solution is quoted from a previous result \citep{Munday2024dbl}.}
    
    \begin{tabular}{l|c|c|c|c}
        Tel / Instr & VLT/UVES & WHT/ISIS & HST/COS & Adopted\\
        \hline
        $T_{\textrm{eff},1}$ [K] & $17230\pm710$ & 16500$^{+400}_{-300}$ & $18630\pm80$ & $17260^{+1380}_{-880}$\\
        $\log$\,g$_1$ [dex]  & $8.408\pm0.027$ & $8.35\pm0.05$ & $8.307\pm0.020$ & $8.350^{+0.066}_{-0.052}$\\
        $M_1$ [$\mathrm{M}_\odot$] & $0.871\pm0.018$ & $0.83\pm0.03$ & $0.810\pm0.013$ & $0.834\pm0.039$ \\
        $T_{\textrm{eff},2}$ [K] & $20190\pm280$ & $20200\pm300$ & $18010\pm70$ & $20000^{+400}_{-2000}$ \\
        $\log$\,g$_2$ [dex] & $8.151\pm0.021$ & $8.16\pm0.04$ & $8.178\pm0.018$ & $8.164^{+0.027}_{-0.030}$ \\
        $M_2$ [$\mathrm{M}_\odot$]  & $0.713\pm0.014$ & $0.72\pm0.03$ & $0.727\pm0.013$ & $0.721\pm0.020$ \\
        $M_{\textrm{T}}$ [$\mathrm{M}_\odot$] & $1.584\pm0.022$ & $1.55\pm0.04$ & $1.537\pm0.018$ & $1.555\pm0.044$\\
    \end{tabular}
    \label{tab:atmosphericFits}
\end{table*}

\begin{figure}
    \centering
    \includegraphics[keepaspectratio, trim={0.25cm 0cm 1.5cm 1.25cm},clip,width=10cm]{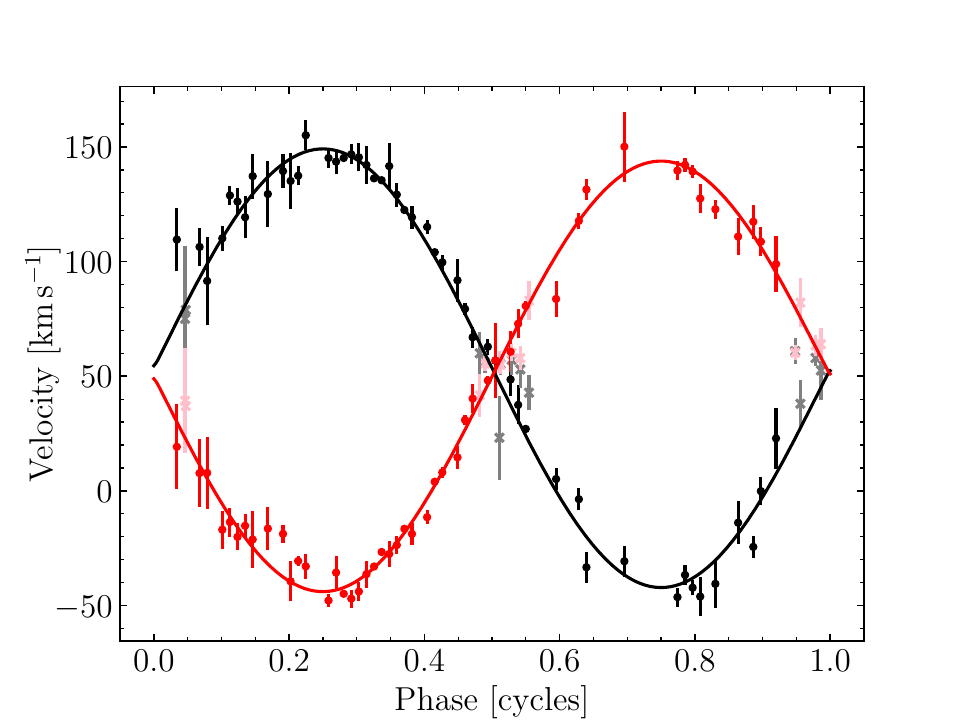}
    \caption{\textbf{The best-fit orbital solution phase-folded on the orbital period.} In black points, the hotter star, and in red the cooler star. The RV curves are plotted showing the velocity of the two stars across a full orbit, binned into 80 evenly spaced phase bins. In faded colours and with crossed markers are the RVs that were masked in searching for an orbital solution, which are also listed in Supplementary Data 1. One-sigma Error bars are given as the standard deviation of 1000 bootstrapping iterations.}
    \label{fig:LSperiodogramRVcurve}
\end{figure}

\section{Results and discussion}\label{sec2}
\subsection{Physical and orbital properties of WDJ181058.67+311940.94}
WDJ181058.67+311940.94 was first discovered as part of the DBL survey \citep{Munday2024dbl} which searches for double-lined double white dwarfs using medium resolution spectra ($R\approx8000$--$9000$). Fits to these identification spectra indicated the source to be a double white dwarf binary with a high total mass. Afterwards, we launched an observational campaign to acquire time-series spectroscopy of the source to confirm the masses derived through the atmospheric parameters and resolve the orbital period. We obtained phase-resolved radial velocities of WDJ181058.67+311940.94 with the following instruments and telescopes: the Intermediate-dispersion Spectrograph and Imaging System (ISIS) on the 4.2\,m William Herschel Telescope; the Intermediate Dispersion Spectrograph (IDS) on the 2.5\,m Isaac Newton Telescope; the FIbre-fed Echelle Spectrograph (FIES) and the NOT Alhambra Faint Object Spectrograph and Camera (ALFOSC) on the 2.56\,m Nordic Optical Telescope; a continuous observing window of 4.5\,hr using the UV-Visual Echelle Spectrograph (UVES) on the 8.2\,m Very Large Telescope (VLT). 

The UVES data was used for an improved accuracy of the atmospherically derived masses from spectral fits because of its full visible coverage. Precise radial velocity measurements of the target were simultaneously obtained and with this an unambiguous determination of the orbital period. As a further test for consistency of the atmospheric solution with a unique dataset, we also fit a two-star solution to a previously published Hubble Space Telescope (HST) Cosmic Origins Spectrograph (COS) spectrum \citep{Sahu2023}. The resultant stellar parameters found by fitting each dataset are quoted in Table~\ref{tab:atmosphericFits} and spectral fits to the optical and ultra-violet data are plotted in Fig.~\ref{fig:SpecPhotFit}. Considering the measurements from all datasets, we find stellar parameters of $T_{\textrm{eff},1} = 17260^{+1380}_{-880}$\,K, $\log$\,g$_1=8.350^{+0.066}_{-0.052}$\,dex, $M_1=0.834\pm0.039$\,$\mathrm{M}_\odot$ for the primary (more massive) star and $T_{\textrm{eff},2}$=$20000^{+400}_{-2000}$\,K, $\log$\,g$_2$=$8.164^{+0.027}_{-0.030}$\,dex, $M_2=0.721\pm0.020$\,$\mathrm{M}_\odot$ for the secondary (less massive) star, leading to a total system mass of $M_{\textrm{T}}=1.555\pm0.044$\,$\mathrm{M}_\odot$. 

All other data was used exclusively for radial velocity measurements at H$\alpha$ to precisely quantify the motion of the stars across all orbital phases and to improve the precision of the period. A Lomb-Scargle periodogram of all radial velocity measurements, which was optimised for physical limits of the system, revealed one clear peak representing the orbital period. The binary parameters are quoted in Table~\ref{tab:allParams} and our phase-folded RV curve with the best-fit orbital solution is depicted in Fig.~\ref{fig:LSperiodogramRVcurve}. The best-fitting orbital parameters are $P=14.23557\pm0.00002$\,hr, $K_1=93.9\pm2.0$\,km\,s$^{-1}$, $K_2=95.7\pm2.1$\,km\,s$^{-1}$, $\gamma_1=50.0\pm1.5$\,km\,s$^{-1}$, $\gamma_2=53.5\pm1.6$\,km\,s$^{-1}$. Being double-lined, the mass ratio is independently solvable without knowledge of the orbital inclination with $q=M_2/M_1=K_1/K_2$, such that the orbitally derived $q=0.98\pm0.03$. This result is in best agreement with the star masses derived from the HST/COS spectrum, which was $q=0.90\pm0.02$. Our derived masses from the VLT/UVES spectra yield a lower $q=0.82\pm0.02$, with the mass of the less massive star being near identical to the ultra-violet, and the adopted value taking into account all measurements indicates a mass ratio of $q=0.86\pm0.04$. The surface gravity of the hotter, less massive star is near identical across fits to all datasets. This is unsurprising given that it contributes more flux than the cooler white dwarf, while its temperature difference between the ultra-violet and optical datasets primarily arises from the fitting of the slope of the spectral energy distribution across the ultra-violet. Forcing the orbitally-derived $q=0.98\pm0.03$ in the atmospheric fit leads the surface gravity of the secondary to increase and the surface gravity of the primary to decrease to fit the broadness of the Balmer line profiles well. The secondary would thus be more massive, the primary less massive, and as such, including for a mass ratio of approximately one, all evidence points towards WDJ181058.67+311940.94 being a super-Chandrasekhar mass double white dwarf. 

The critical time at which the two stars reach closest approach is calculable using \citep{Peters1964GravRadTwoPointMasses}
\begin{equation}
    T_c(a_0) =\frac{5}{256} \frac{a_0^4 c^5}{G^3 M_1 M_2 (M_1+M_2)}
\end{equation}
where $a_0$ is the semi-major axis of the binary at present day and, for WDJ181058.67+311940.94, $a_0=0.01601\pm0.00015$\,AU. This indicates that the stars will come into contact in $22.6\pm1.0$\,Gyr, while the less-massive component will begin Roche lobe overflow and initiate mass transfer approximately 100\,yrs before the demise of binary.

\begin{table*}
    \centering
    \caption{\textbf{Positional, atmospheric and orbital parameters for WDJ181058.67+311940.94.} The primary and secondary stars correspond to the more massive and less massive components, respectively. The temperatures, surface gravities and masses quoted are the adopted values from the spectroscopic fits, which were determined considering data from all sources (see Table~\ref{tab:atmosphericFits}).}
    \begin{tabular}{l|l|l|l}
        Parameter & Unit & Value & Uncertainty \\
        \hline
        Right ascension & deg (2016) & 272.744360834 & $\pm$0.000000005\\ % mas turned to deg
        Declination & deg (2016) & 31.327961071 & $\pm$0.000000005\\ % mas turned to deg
        Reference epoch & HJD (UTC) & 2458587.6663 & $\pm$0.0018 \\
        Orbital period & day & 0.5931479 & $\pm$0.0000009\\
        Gaia parallax & mas & 20.438 & $\pm0.023$\\
        Fitted parallax & mas & 20.402 & $\pm0.003$\\
        Primary temperature & K & 17\,260 & ${+1380}/{-880}$ \\
        Secondary temperature & K & 20\,000 & ${+400}/{-2000}$ \\
        Primary surface gravity & dex & 8.350 & ${+0.066}/{-0.052}$ \\
        Secondary surface gravity & dex & 8.164 & ${+0.027}/{-0.030}$ \\
        Primary mass & M\(_\odot\) & 0.834 & $\pm0.039$\\
        Secondary mass & M\(_\odot\) & 0.721 & $\pm0.020$\\
        System mass & M\(_\odot\) & 1.555 & $\pm0.044$\\
        Primary semi-amplitude & km\,s$^{-1}$ & 93.9 & $\pm$2.0\\
        Secondary semi-amplitude & km\,s$^{-1}$ & 95.7 & $\pm$2.1\\
        Primary velocity offset & km\,s$^{-1}$ & 50.0 & $\pm1.5$\\
        Secondary velocity offset & km\,s$^{-1}$ & 53.5 & $\pm1.6$\\
        Merger time & Gyr & 22.6 & $\pm1.0$\\
    \end{tabular}
    \label{tab:allParams}
\end{table*}

\subsection{Modelling the fate of the binary system}
\label{sec:SupernovaModelling}
\begin{figure}[H]
    \centering
    \includegraphics[trim={0.25cm 0cm 0.15cm 0cm},width=\textwidth]{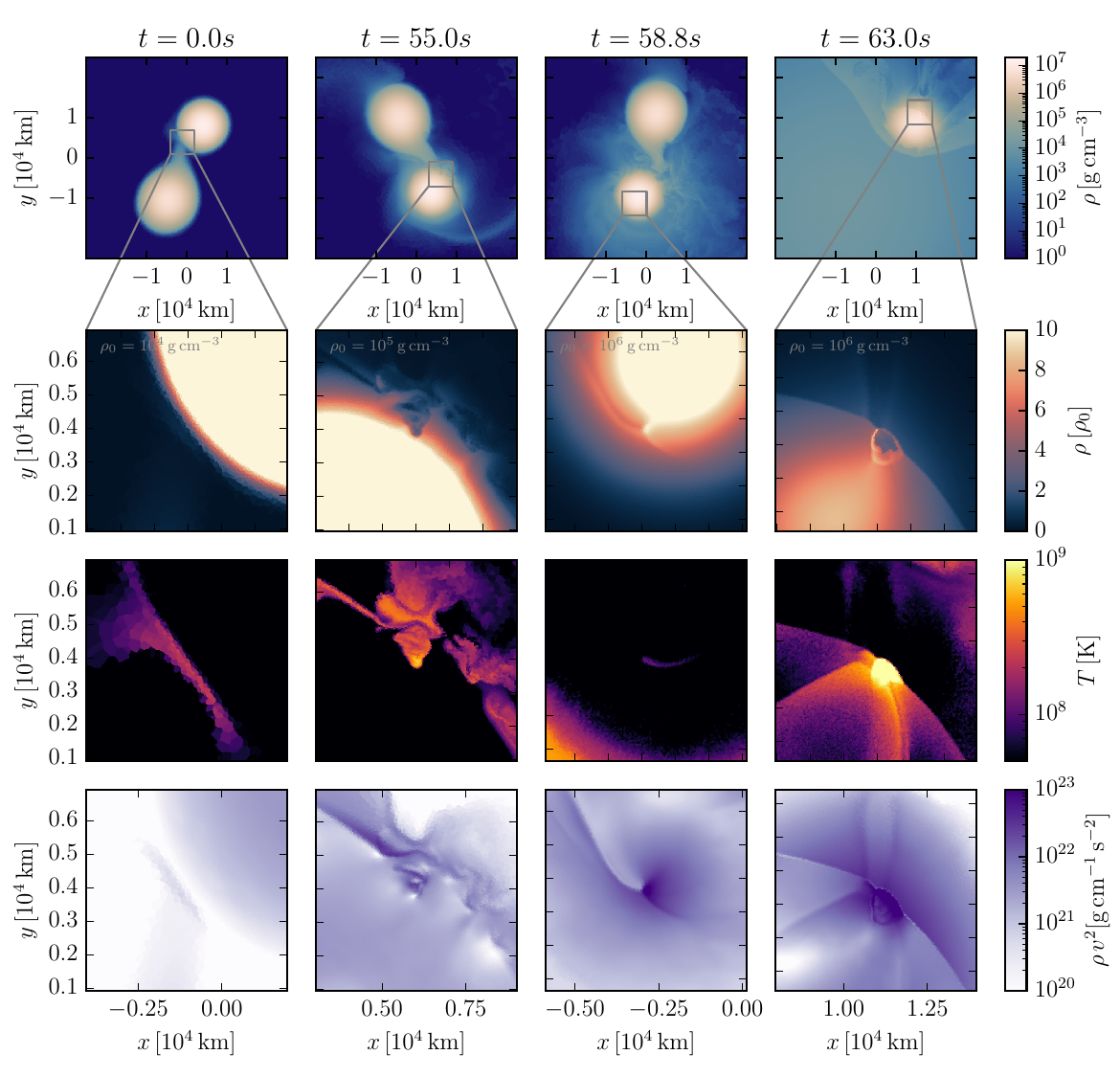}
    \caption{\textbf{Time evolution in slices of the binary systems close to merger.} The first column shows the time when we stop the accelerated inspiral and continue to evolve the binary system self-consistently. The second and third columns show the time when the helium detonation ignites on the surface of the primary white dwarf, and the time when the shock wave that is driven into the core of the primary by the helium detonation converges in a single point. The fourth column shows the same shock convergence in the core of the secondary white dwarf. The top row shows slices of density in the plane of rotation and the three below are zoomed insets at the point of interest. From top to bottom: density, temperature, and kinetic energy density. The shock convergence points in both white dwarfs occur at densities high enough to very likely ignite a carbon detonation and destroy the white dwarf.}
    \label{fig:MergerEvolution}
\end{figure}

\begin{figure}[H]
    \centering
    \includegraphics[width=\textwidth,trim={0.25cm 0cm 0.5cm 0cm}]{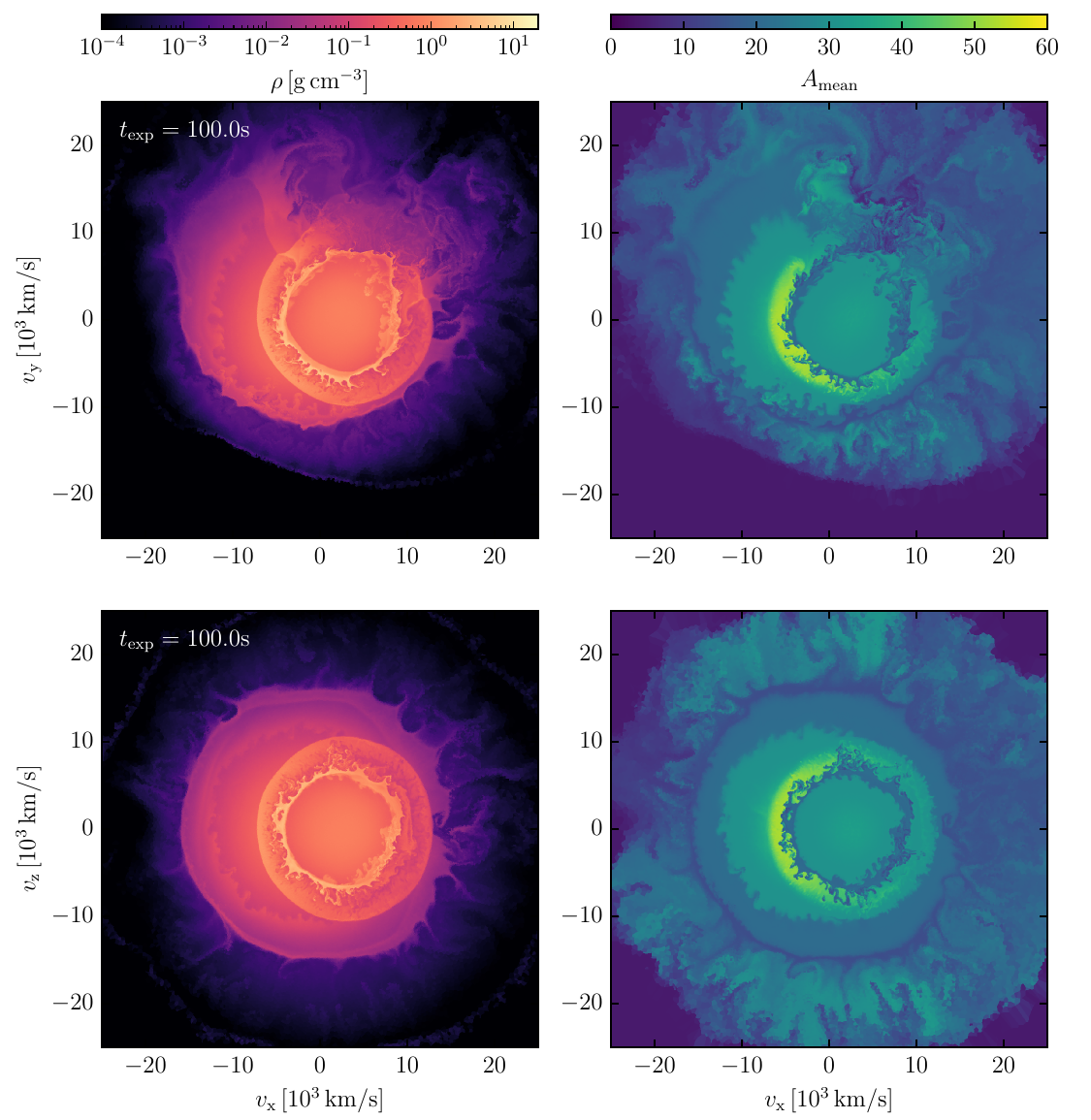}
    \caption{\textbf{Ejecta profiles after the explosion of both stars.} Slices of density (left column) and mean atomic weight (right column) of the supernova ejecta in homologous expansion $100\,\mathrm{s}$ after ignition of the first helium detonation are included. The top row shows slices in the original plane of rotation, the bottom row slices perpendicular to it. The outer layers are close to spherically symmetric, though significant deviations from spherical symmetry exist in the plane of rotation. The iron group elements (including $^{56}\mathrm{Ni}$) are essentially all produced in the explosion of the primary white dwarf and form a half-sphere around the ejecta of the secondary white dwarf.}
    \label{fig:Ejecta}
\end{figure}

To understand the fate of the binary system, we simulate its interaction when it is just about to merge using the star masses obtained through spectral fits to the VLT/UVES data. A movie of this simulation is presented in the supplementary material. The spectral fits indicate that carbon-oxygen cores are appropriate for both white dwarfs, so we consider it highly unlikely that an accretion induced collapse will occur. This would require the more massive white dwarf to have a mass higher than $1.2$\,M$_\odot$ with an oxygen-neon core to avoid dynamical ignition before or during the merger.  We use the moving-mesh code \textsc{arepo} \citep{Arepo,Pakmor2016,Weinberger2020} in a similar setup as prior work \citep{Pakmor2022}. The stars were given realistic composition profiles and placed in co-rotation before applying an accelerated inspiral term that removes angular momentum in the same way as gravitational waves. We switch on a live nuclear reaction network with $55$ isotopes \citep{Pakmor2012,Pakmor2022} at an orbital period of $39\,\mathrm{s}$ as the temperature of the accretion stream at the impact spot approaches that required for a thermonuclear runaway. We show an overview of the dynamic evolution of the binary system in Fig.~\ref{fig:MergerEvolution}. 
The interaction of the accretion stream with the surface of the primary white dwarf ignites a helium detonation close to the point of interaction (second column of Fig.~\ref{fig:MergerEvolution}).
The helium detonation then wraps around the primary white dwarf and sends a shock wave into its core that converges at a single point. 
This ignites a second detonation that completely destroys the primary white dwarf. When the shock wave of its explosion hits the secondary white dwarf, the double detonation mechanism repeats itself. The shock wave from the detonation of the primary ignites a helium detonation near the surface of the secondary that drives a shock wave into its core. It is sufficient to ignite the core detonation, destroying the secondary white dwarf as well.

There is no bound remnant and the ejecta of the explosion contain the total mass of the initial binary, having a total explosion energy of $1.2\times10^{51}\,\mathrm{erg}$. We show the structure and composition of the ejecta in Fig.~\ref{fig:Ejecta}. The outermost layers of ejecta are the ashes of the helium detonation of the primary white dwarf. They consist mostly of intermediate mass elements, dominated by silicon, sulphur, and argon. Below them sit the ashes of the carbon-oxygen core of the primary white dwarf. They again consist mostly of intermediate mass elements, but also contain $0.13\,\mathrm{M_\odot}$ of iron group elements, in particular $0.10\,\mathrm{M_\odot}$ of radioactive $^{56}\mathrm{Ni}$ that will power the lightcurve. The resulting supernova has a maximum brightness in the $B$-band of $M_\mathrm{B}=-16.4$ ($m_B=-14.7$), a maximum brightness in the $V$-band of $M_\mathrm{V}=-17.8$ ($m_V=-16.1$) and will most likely appear as a subluminous type Ia supernova.

\subsection{Galactic rates of super-Chandrasekhar mass double white dwarfs}
We can use WDJ181058.67+311940.94 to observationally predict the number of super-Chandrasekhar mass double white dwarfs in the Milky Way. We start by assuming that WDJ181058.67+311940.94 and NLTT~12758 are the only two within 49\,pc and make the rudimentary assumption that double white dwarfs are evenly scattered around the Milky Way having a cylindrical disk with radius $R_\textrm{max} = 15$\,kpc and scale height $h_z = 300$\,pc. The white dwarf birthrate is estimated to be $\approx1.4\times10^{-12}$\,pc$^{-3}$\,yr$^{-1}$ \citep{Holberg2016}, there are 1076 white dwarfs within the volume complete 40\,pc \textit{Gaia} sample \citep{OBien2024} and extrapolated to 49\,pc we would have 1978 white dwarfs. This means that the birthrate of super-Chandrasekhar mass double white dwarfs in the Galaxy becomes greater than approximately $6.0\times10^{-4}$\,yr$^{-1}$.

Moreover, we can also calculate an observed rate of type Ia supernovae arising from super-Chandrasekhar mass double white dwarfs using WDJ181058.67+311940.94 ($T_c=22.6\pm1.0$\,Gyr) and NLTT~12758 ($T_c=139\pm9$\,Gyr). The frequency of the two events combined imply a supernova rate of about once every 19\,Gyr within 49\,pc, or $\left(1.04\pm0.04\right)\times10^{-16}$\,yr$^{-1}$\,pc$^{-3}$. When fully extrapolated with the cylindrical disk approximation, the observed rate of type Ia supernovae from super-Chandrasekhar mass double white dwarfs in the Milky Way hence becomes at least $\left(4.4\pm0.2\right)\times10^{-5}$\,yr$^{-1}$, though the quoted uncertainty does not account for uncertainties on the Galactic model. This result serves as a minimum based on the 49\,pc population as it remains possible that other systems exists within the same radius.

Evidently, the magnitude of super-Chandrasekhar mass systems approaches the $\left(2.8\pm0.6\right)\times10^{-3}$\,yr$^{-1}$ rate predicted for all evolutionary channels leading to a type Ia \citep{LiSupernovaeRates2011, Maoz2014Type1aProgenitors, Liu2023supernovaReview}, but we must recall that these two systems are set to come together in over a Hubble time and consider that the present observed supernova rate from these systems is about sixty times smaller. As such, the rates from Milky Way progenitors through the hot subdwarf binary channel and the double white dwarf channel are about the same, together accounting for about 3\% of the Galactic rate. Synthetic populations suspect that around 60\% of the Galactic birthrate of type Ia progenitors comes from the double degenerate channel \citep{2010SCPMA..53..586W, 2018MNRAS.473.5352L}, like is the case for WDJ181058.67+311940.94. The large missing fraction is especially mysterious given the high completion rate of the 40\,pc sample of white dwarfs \citep{OBien2024}. Contribution to the double white dwarf type Ia rate from sub-Chandrasekhar mass limits detonation could at least be a partial solution to make up for the deficit, where a mass-period distribution of double white dwarfs in a volume/magnitude limited sample serves as a means to put this to the test \citep{Munday2024dbl}. To date, there have been no sub-Chandrasekhar mass type Ia progenitor candidates inside of a 50\,pc radius, such that ongoing efforts are crucial to properly quantify the number of massive double white dwarf binaries in our local neighbourhood and the Milky Way.

\section{Conclusion}\label{sec13}
We have presented the first confirmed to be compact, super-Chandrasekhar mass double white dwarf binary which will merge in close to a Hubble time, having an orbital period of 14.24\,hr. With a total mass of $1.555\pm0.044$\,\(\textup{M}_\odot\), WDJ181058.67+311940.94 is the most massive double white dwarf binary confirmed to date. %, followed by one other super-Chandrasekhar mass double white dwarf having a merger time of approximately 137\,Gyr \citep{Kawka2017}. 
We predict it to explode as a quadruple detonation and be destroyed completely. With all the mass ejected and a total explosion energy of $1.2\times10^{51}\,\mathrm{erg}$, but only $0.1\,\mathrm{M_\odot}$ of $^{56}\mathrm{Ni}$ in the ejecta, it will appear as a subluminous type \Romannum{1}a supernova with a peak apparent magnitude of approximately $m_B=-14.7$ and $m_V=-16.1$.

The lack of observational evidence of compact and massive double white dwarf binaries has long troubled the theory that double white dwarfs are the dominating evolutionary channel of type \Romannum{1}a detonations \citep{Maoz2012supernovaReview}. WDJ181058.67+311940.94 provides tentative observational evidence that super-Chandrasekhar mass systems with short merger times do exist in the Milky Way, and when combined with the close proximity of 49\,pc the rate of super-Chandrasekhar mass double white dwarfs born in the Milky Way is at least $6.0\times10^{-4}$\,yr${^{-1}}$. This draws closer the disparity between the observed and predicted birthrates of super-Chandrasekhar mass systems, though the observed rate is still approximately twice smaller. However, there remains a large deficit in the rate of type Ia supernovae from the progenitor systems. A small fraction of the Milky Way rate is accounted for, now with an equal contribution from double white dwarf and white dwarf+hot subdwarf binaries.

Being discovered through a medium-resolution search of overluminous double white dwarfs \citep{Munday2024dbl}, which up to a magnitude limit of $G<17$\,mag is approximately $20\%$ complete, it is entirely plausible that more super-Chandrasekhar mass double white dwarfs reside in our Galactic neighbourhood and that we have the spectroscopic ability to resolve the formation channel of type \Romannum{1}a supernovae. Deeper completeness through photometric  and spectroscopic surveys in the coming years, as well the inauguration of space-based gravitational wave detectors in the next decade, will be pivotal in detecting ultra-compact binaries on the cusp of detonation \citep{Korol2017prospects, 2024arXiv240703935K}. Combined efforts surveying type \Romannum{1}a progenitors across the full range of orbital periods will be the ultimate means to accurately quantify the contribution of double white dwarfs to type \Romannum{1}a supernovae.

\newpage
\section{Methods}\label{sec11}
\subsection{Observations}
WDJ181058.67+311940.94 was first discovered as part of the DBL survey \citep{Munday2024dbl} using medium-resolution spectra ($R=8\,800$) on the 4.2\,m William Herschel Telescope with the Intermediate-dispersion Spectrograph and Imaging System (ISIS). Two other ISIS exposures were taken on the nights 13 \& 14 April 2019 using the R600B and R1200R gratings with a 1.2$^{\prime\prime}$ slit resulting in a spectral resolution of $R=3\,000$ at H$\alpha$ and these spectra are included in the full orbital analysis of the double white dwarf. The blue and red setups had a wavelength calibration accuracy of approximately 3\,km\,s$^{-1}$ and 2\,km\,s$^{-1}$, respectively.

We conducted a continued observational campaign to derive phase-resolved radial velocities (RVs) of the double white dwarf binary. We utilised the 2.5\,m Isaac Newton Telescope (INT) with the Intermediate Dispersion Spectrograph (IDS) over the nights 4-7 September 2019 (11 exposures, 1800\,s each), and 24 September 2019 (4 exposures, 900\,s each) with the Red+2 detector and a 1.2$^{\prime\prime}$ slit width, resulting in a spectral resolution of $R=6\,300$. Further phase-resolved spectra were taken with the INT on the nights 25 \& 26 August 2024 with the H1800V grating at a resolution of $R=9\,400$ (20 exposures, 1500\,s each). An arc lamp exposure was taken every 45\,min of observing time and the science images were wavelength calibrated by interpolation of the nearest two arcs. The wavelength calibration accuracy per frame was approximately 2\,km\,s${^{-1}}$.

Bias, flat field and spectrophotometric flux standard star images were taken on all nights and applied in the reduction. All data from the WHT and the INT were reduced using the \textsc{molly} suite \citep{Marsh2019Molly} using an optimal extraction algorithm \citep{Marsh1989optimalExtraction}.

These data were supplemented with 18 exposures of length 1\,500\,s on the 2.56\,m Nordic Optical Telescope (NOT) using the FIbre-fed Echelle Spectrograph (FIES) in low-resolution mode ($R=25\,000$), having a wavelength calibration accuracy of approximately $\pm150\,$m\,s$^{-1}$. Observations were obtained through a staff queue at random times, typically being two consecutive exposures, and through a NOT fast-track proposal. All FIES data were reduced using its automated data reduction pipeline -- FIEStool \citep{FIEStool}. We also obtained 5 exposures with the NOT Alhambra Faint Object Spectrograph and Camera (ALFOSC) with a 0.5$^{\prime\prime}$ slit width, producing spectra at R=10\,000 with wavelength range 6330-6870\,\AA~on 1 \& 2 June 2024. The data were reduced with the \textsc{pypeit} python package \citep{Prochaska2020pypeit}.

A continuous observing window of 4.5\,hr was obtained through directors discretionary time on the 8.2\,m Very Large Telescope (VLT) with the UV-Visual Echelle Spectrograph (UVES). Each exposure lasted for 730\,s with a readout time between exposures of 45\,s, totalling 20 exposures. We employed an observing setup of the dichroic 1 mode with central wavelengths of 3900\,\AA~and 5640\,\AA~for the blue and red arm, giving a wavelength range that covered the full visible spectrum besides gaps of 80\,\AA~at 4580\,\AA~and 5640\,\AA. A slit width of 1.0'' and a 2$\times$2 binning granted a spectral resolution $R=20\,000$ and the wavelength calibration accuracy was approximately 200\,m\,s$^{-1}$ \citep{WhitmoreUVES2010, WhitmoreUVES2015}.

In deriving final RV errors for these data (Supplementary Data 1), the wavelength calibration error was added in quadrature to the statistical error.

\subsection{Atmospheric fitting of optical data}
We used the package WD-BASS \citep{Munday2024wdbass} to fit atmospheric parameters to the spectra from VLT/UVES. For synthetic spectra, we utilised the 3D-NLTE model grid introduced in \cite{Munday2024dbl}, which was constructed using the 3D-LTE models of \cite{Tremblay2015} with a further NLTE correction factor applied using the NLTE and LTE synthetic spectra described in \cite{Kilic2021HiddenInPlainSight}. The two stars were scaled using temperature-$\log~g$-radius relationships with the evolutionary track models of \cite{Istrate2016} when $M\leq0.393$\,M$_\odot$, \cite{Althaus2013} when $0.393<M<0.45\,\mathrm{M}_\odot$ and the hydrogen-rich envelope evolutionary sequences of \cite{Bedard2020} otherwise. These boundaries come from the expectation that a white dwarf with a mass below $0.45\,\mathrm{M}_\odot$ has a helium core and those larger have a carbon-oxygen core. The model spectra were converted from an Eddington flux to that observed at Earth and reddened with $A(V)=0.0312$\,mag \citep{Lallement2022}, $E(B-V)=A(V)/3.1$ using the reddening curves of \cite{Gordon2023}.

We applied an atmospheric fitting technique that is very similar to that described in \cite{Munday2024dbl} by linearly normalising and fitting the Balmer spectral lines of the UVES data using a Markov Chain Monte Carlo algorithm, maximising the likelihood for a best-fit solution. We also utilised Pan-STARRS photometry \citep{Panstarrs} to perform a hybrid fit using both datasets simultaneously. With high signal-to-noise ratio data, we were able to fit all Balmer lines from H$\alpha$ to H11. Furthermore, to give the photometric and spectroscopic data a similar weight, we applied an extra weighting ($\times$1000) to the photometric fit. Without this weighting, the spectra would have over-dominated the best-fit solution. Only spectra taken at the times where a distinct double-line splitting is evident at H$\alpha$ were fit to avoid fitting degeneracies between the two stars, which were 10 of them (total of 20). In deriving errors, we individually fit each red-arm spectrum that reveals a double-lined H$\alpha$ split along with the nearest-in-time blue-arm spectrum while weighting the photometry by 100$\times$. Then, we took the standard deviation of all measurements to be the error in star's surface gravity and temperature. The new best-fit atmospheric parameters are stated in Table~\ref{tab:atmosphericFits}, which are entirely consistent with previous values \citep{Munday2024dbl}.

\subsection{Atmospheric fitting of ultra-violet data}
\label{subsec:AtmosphericUV}
We performed an independent spectroscopic fit using a previously published Hubble Space Telescope spectrum \citep{Sahu2023}. WDJ181058.67+311940.94 was observed for a single 1\,000\,s exposure using the Cosmic Origins Spectrograph on the 19th February 2022% with the centre of exposure at 05:00~UT
. The observation had a central wavelength of 1291\,\AA~with the G130M grating, giving a resolution of $R=12\,000$--$16\,000$ and a wavelength range of 1130--1430\,\AA~with a gap at 1278–1288\,\AA~due to the positioning of the two detector segments. Given the vastly different method and the fact that WDJ181058.67+311940.94 is not double-lined at Lyman-$\alpha$ in the ultra-violet data, no RVs were extracted, but the predicted RVs of the two stars at the centre of exposure ($-37.8$\,km\,s$^{-1}$ for the more massive and 139.6\,km\,s$^{-1}$ for the less massive star, respectively) were fixed in the fitting procedure.

%https://mast.stsci.edu/search/ui/#/hst/results?resolve=true&target=272.74436083390964%2031.327961071239763&observations=S&radius=10&radius_units=arcseconds&useStore=false&search_key=9eafcb09ab692

Our spectral fitting method is identical to that presented in \cite{Sahu2023} with the only exceptions being that a second hydrogen-rich atmosphere white dwarf is included in the model, that we adopt $A(V)=0.0312$\,mag and that the mid-exposure RV of the two stars is considered. A hybrid (spectroscopic and photometric) fit was performed with no extra error weighting applied using the HST/COS spectrum and photometry from Pan-STARRS $g, r, i, z, y$ \citep{Panstarrs}, fixing the distance to  \textit{Gaia}~DR3 parallax. Updated model atmospheres \citep{Koester2010} with a white dwarf mass-radius relation \citep{Bedard2020} were used to fit the absolute fluxes. Additionally, strong absorption lines affecting the continuum were masked in the COS spectrum \citep[][]{Sahu2023}. To address the inconsistencies reported between ultra-violet and optical parameters \citep{Sahu2023}, a systematic offset of 1\% in $T_\textrm{eff}$ and 0.1\,dex in $\log(\textrm{g})$ were added to the ultra-violet values of both stars in the hybrid fitting, while trial values in the optical were unchanged. The best-fit model to the spectra are shown in Fig.\ref{fig:SpecPhotFit}, and the results of our atmospheric fitting in Table~\ref{tab:atmosphericFits} with comparison to the optical solution. We found a total mass of $1.537\pm0.018$\,M$_\odot$ through this analysis, which again is consistent with previous values \citep{Munday2024dbl}.

To provide a final adopted value from the atmospheric fitting inclusive of the results from the optical and the ultra-violet datasets, we concatenated the distributions obtained for each parameter to then quote the median and 68\% confidence interval on the T$_\textrm{eff}$ and $\log(g)$, while interpolating these parameters to find obtain masses. The adopted values are quoted in Tables~\ref{tab:atmosphericFits} and Table~\ref{tab:allParams}.

\subsection{Radial velocities and orbital parameters}
WD-BASS \citep{Munday2024wdbass} was again used to obtain radial velocities (RVs) for all of the optical spectra. The best-fit synthetic spectrum agrees with the data extremely well (see Fig.~\ref{fig:SpecPhotFit}), but even with the correction of NLTE effects to the model grid line cores, the synthetic model flux is over-predicted in the line cores of H$\alpha$. To obtain the most accurate template for RV extraction possible, we fit a Gaussian model to the H$\alpha$ line cores of both stars combined with a 4-term polynomial to model the broader wings of H$\alpha$, all within 10\,\AA~of the H$\alpha$ centre. The centre of a H$\alpha$ absorption was isolated as the splitting of the two stars is most apparent around the non-thermal equilibrium line cores and hence the stars are most easily disentanglable. This method best modelled the shape of the spectral area around the line cores for the high signal-to-noise ratio and high resolution UVES spectra, but not for all other data sources. Instead, we took the result of the best-fit synthetic spectrum and added an extra Gaussian component at the line cores of H$\alpha$ for both stars \cite[following the method described in Section~4.4 of][]{Munday2024dbl}, which improved the line-core shape significantly. The Gaussians were fit to all relevant spectra simultaneously and this final template spectrum was then used for RV extraction in WD-BASS. We started by fitting the RV of both stars to each spectrum by taking the median of 1000 bootstrapping iterations and taking errors as the standard deviation of this bootstrapped posterior distribution.

With the full set of 82 RV measurements (Supplementary Data 1), we then searched for an orbital period, $P$, by minimising the $\chi^2$ of equation~\ref{eqn:massFunction} for trial semi-amplitudes ($K_1$, $K_2$) and velocity offsets ($\gamma_1$, $\gamma_2$) of each star using a least squares algorithm, where
\begin{equation}
    P K_{\textrm{1,2}}^3 = \frac{2\pi GM_{\textrm{2,1}}^3 \sin(i)^3}{(M_{\textrm{1,2}} + M_{\textrm{2,1}})^2}
    \label{eqn:massFunction}
\end{equation}
An upper bound on the semi-amplitude, $K_{\textrm{max}, 1}$ and $K_{\textrm{max}, 2}$ was set for a trial period set by applying an edge-on ($i=90^\circ$) inclination for a 1.4\,$\mathrm{M}_\odot + 0.15\,\mathrm{M}_\odot$ double white dwarf in a Keplerian orbit (the maximum and observed minimum mass of a white dwarf, respectively). There is no indication of eccentricity from the RVs, so the orbit is assumed to be circularised ($e=0$). In the process, we noticed a deviation from Keplerian motion around conjunction which is caused by degeneracy in the fitted RVs as the stars spectrally overlap. This is unsurprising as the velocity resolution of the ALFOSC, ISIS and IDS data was around 30--40\,km\,s$^{-1}$, while in the higher resolution FIES spectra a lower signal-to-noise ratio led to the same degeneracies. We decided to ignore these RVs when fitting the orbital motion by masking measurements which are within 15\,km\,s$^{-1}$ of the RV of each star at conjunction. All RVs from the UVES spectra within this range were utilised as its high signal-to-noise combined with twice the velocity resolution did not cause any noticeable deviation. With the final periodogram, two prominent peaks appear at very similar solutions; the solution that we adopt with a 14.2356\,hr period and another at 14.2308\,hr, but the second one can be rejected owing to a gravitational redshift difference that would be a strong outlier from that expected in the atmospheric solution.

Returning to equation~\ref{eqn:massFunction} with the final orbital solution and taking into account all combinations of masses from the atmospheric analysis, we conclude that WDJ181058.67+311940.94 has an inclination $i\approx 35-45$\,deg. We analysed the TESS \citep{Ricker2015TESS} light curve of WDJ181058.67+311940.94 in all cadences to search for any photometric signature of photometric variability with Lomb-Scargle \citep{Lomb1976, Scargle1982} and boxed-least-squares periodograms but found no variation on the orbital period. For an eclipse to be witnessed in this system, the inclination would have to be above 89.64\,deg and photometric variability from ellipsoidal modulation or irradiation is minute for a system with 14.24\,hr orbital period. The Doppler beaming from the two stars is nullified by their opposing motion of near-identical RV amplitudes and a similar flux contribution \citep{Hermes2014}, hence non-eclipsing forms of variability are not expected.

\subsection{Modelling the fate of the binary system}
We created two white dwarfs from the pre-main sequence phase using the stellar evolution code \textsc{mesa} \citep{Paxton2011, Paxton2013, Paxton2015, Paxton2018, Paxton2019, Jermyn2023}, evolving them to carbon-oxygen white dwarfs of $0.87\,\mathrm{M_\odot}$ and $0.71\,\mathrm{M_\odot}$. These masses align with observations from VLT/UVES spectra fitting. Compared to previous merger simulations, using self-consistent models evolved in \textsc{mesa} allows us to start from realistic composition profiles. In particular, the two white dwarfs have a helium shell of $8\times10^{-3}\,\mathrm{M_\odot}$ (for the $0.71\,\mathrm{M_\odot}$ white dwarf) and $3\times10^{-3}\,\mathrm{M_\odot}$ (for the $0.87\,\mathrm{M_\odot}$ white dwarf), respectively.

We then created two 3D white dwarfs in hydrostatic equilibrium with the same masses and abundance profiles in \textsc{arepo}. We resolved the white dwarfs with cells with a roughly constant mass of $10^{-7}\,\mathrm{M_\odot}$ and used a passive scalar to resolve the helium shells of both white dwarfs even better with a mass resolution of $10^{-8}\,\mathrm{M_\odot}$. We relaxed both white dwarfs in isolation for $10$ dynamical timescales, actively dampening any gas velocities for the first half of this time. The density and composition profiles of the relaxed white dwarfs, in particular close to the surface, well resembled the initial 1D profiles obtained from \textsc{mesa}.

We put both white dwarfs into a binary system in co-rotation with an initial period of $73\,\mathrm{s}$. At this period the separation is about $1.5$ times larger than the separation where the secondary white dwarf will fill its Roche-lobe. We apply an accelerated inspiral term that removes angular momentum in the same way as gravitational waves, but on a much faster timescale. This way we obtain a binary system in equilibrium when mass transfer starts on a scale that we can resolve in the simulation. At this time, the physical system will have transferred mass at a low rate for possibly hundreds of years, but the total mass transferred is likely negligible. The secondary white dwarf eventually starts filling and then overfilling its Roche-lobe, and we stopped the accelerated inspiral when the density at the inner Lagrange point between the white dwarfs reached $2\times10^4\,\mathrm{g\,cm^{-3}}$. Only then the density in the accretion stream becomes large enough to dynamically affect the surface of the primary white dwarf \citep{Guillochon2010,Pakmor2013,Pakmor2022}.

The binary system has now shrunk to a separation of $0.03\,\mathrm{R_\odot}$ and an orbital period of $39\,\mathrm{s}$. We then continued to evolve the binary system conservatively and switched on a live nuclear reaction network with $55$ isotopes \citep{Pakmor2012,Pakmor2022}. After evolving the binary system conservatively for $55\,\mathrm{s}$, the interaction of the accretion stream with the surface of the primary white dwarf ignites a helium detonation close to the point of interaction (second column of Fig.~\ref{fig:MergerEvolution}), consistent with previous simulations of more massive white dwarf binaries \citep{Pakmor2013,Pakmor2021,Pakmor2024}. As in the classic double detonation scenario where the helium detonation is caused by instabilities in a massive helium shell \citep{Livne1990,Fink2010}, the helium detonation wraps around the primary white dwarf. It sends a shock wave into the core of the white dwarf, that converges in a single point at a density of $9.6\times10^6\,\mathrm{g\,cm^{-3}}$. Because of a lack of numerical resolution, the simulation does not self-consistently ignite a carbon detonation there, but resolved ignition simulations indicate that at this density we expect a detonation to form at the convergence point \citep{Seitenzahl2009,Shen2014}. To model the ignition of the detonation when the shock converges in the simulation, we set the temperature of $178$ cells that contain $1.8\times10^{-5}$\,M$_\odot$ around the convergence point to $5\times10^9\,\mathrm{K}$. This injects $4.8\times10^{46}\,\mathrm{erg}$ (negligible compared to the energy release of the whole simulation) and ignites the detonation. The detonation completely destroys the primary white dwarf. When the shock wave of its explosion hits the secondary white dwarf, the double detonation mechanism repeats itself. The shock wave ignites a helium detonation that drives a shock wave into the core and converges at a density of $8.5\times10^6\,\mathrm{g\,cm^{-3}}$. In this case, carbon burning even starts at the convergence point, but not strongly enough to start a detonation. We again ignite a detonation at the convergence point by setting the temperature of $708$ cells that contain $6.9\times10^{-5}$\,M$_\odot$ to $6\times10^9\,\mathrm{K}$, which injects $8.2\times10^{47}\,\mathrm{erg}$ and is sufficient to ignite the detonation that then destroys the secondary white dwarf as well.

The total explosion energy is $1.2\times10^{51}\,\mathrm{erg}$. The core of the secondary white dwarf ignites $4.2\,\mathrm{s}$ after the core of the primary white dwarf. At this time the ashes of the primary white dwarf have already expanded far beyond the secondary white dwarf. So when the latter explodes as well, its ejecta expand into and remain in the centre of the ejecta of the primary white dwarf \citep{Pakmor2022}. The outermost layers of ejecta are the ashes of the helium detonation of the primary white dwarf. The centre of the ejecta consists of the ashes of the secondary white dwarf, which contain $0.25\,\mathrm{M_\odot}$ of oxygen, $0.4\,\mathrm{M_\odot}$ of intermediate mass elements, and only $0.01\,\mathrm{M_\odot}$ of iron group elements with a roughly equal fraction of $^{56}\mathrm{Ni}$ and $^{54}\mathrm{Fe}$.

We obtained preliminary synthetic light curves from spherically averaging the ejecta and computing light curves with the Monte-Carlo radiation transport code \textsc{artis} \citep{Kromer2009,Sim2007}. The resulting supernova has a maximum brightness in the B-band of $M_\mathrm{B}=-16.4$ ($m_B=-14.7$) and a maximum brightness in the V-band of $M_\mathrm{V}=-17.8$ ($m_V=-16.1$), consistent with traditional double detonation models of single white dwarfs with a similar mass as our primary white dwarf, because the secondary white dwarf does not produce any significant amount of radioactive $^{56}\,\mathrm{Ni}$ \citep{Sim2010,Shen2021,Collins2022}. That said, our explosion likely avoids the imprint of thick helium shells on light curves and spectra \citep{Kromer2010,Collins2022,Collins2023}. It will most likely appear as a subluminous type Ia supernova. However, the obvious large-scale asymmetries visible in Fig.~\ref{fig:Ejecta} indicate that 3D synthetic observables will be needed to make any reliable statement about the expected display of this supernova \citep{Collins2022,Pakmor2024}. They will be presented and discussed as part of a larger sample of merger simulations in the future. This new simulation also supports previous work which suggests that both stars will explode in massive double white dwarf binaries that are about to merge \citep{Pakmor2022,Boos2024,Shen2024}.

\bmhead{Acknowledgements}
We thank Stephan Geier for their insightful comments during the study. JM was supported by funding from a Science and Technology Facilities Council (STFC) studentship. IP acknowledges support from The Royal Society through a University Research Fellowship (URF/R1/231496). DJ acknowledges support from the Agencia Estatal de Investigaci\'on del Ministerio de Ciencia, Innovaci\'on y Universidades (MCIU/AEI) and the European Regional Development Fund (ERDF) with reference PID-2022-136653NA-I00 (DOI:10.13039/501100011033). DJ also acknowledges support from the Agencia Estatal de Investigaci\'on del Ministerio de Ciencia, Innovaci\'on y Universidades (MCIU/AEI) and the the European Union NextGenerationEU/PRTR with reference CNS2023-143910 (DOI:10.13039/501100011033). This research received funding from the European Research Council under the European Union’s Horizon 2020 research and innovation programme number 101002408 (MOS100PC). ST acknowledges support from the Netherlands Research Council NWO (VIDI 203.061 grant). AB is a Postdoctoral Fellow of the Natural Sciences and Engineering Research Council (NSERC) of Canada. Based on observations collected at the European Organisation for Astronomical Research in the Southern Hemisphere under ESO programme 113.27QU. The Isaac Newton Telescope and the William Herschel Telescope are operated on the island of La Palma by the Isaac Newton Group of Telescopes in the Spanish Observatorio del Roque de los Muchachos of the Instituto de Astrofísica de Canarias. Based on observations made with the Nordic Optical Telescope, owned in collaboration by the University of Turku and Aarhus University, and operated jointly by Aarhus University, the University of Turku and the University of Oslo, representing Denmark, Finland and Norway, the University of Iceland and Stockholm University at the Observatorio del Roque de los Muchachos, La Palma, Spain, of the Instituto de Astrof\'isica de Canarias. The data presented here were obtained in part with ALFOSC, which is provided by the Instituto de Astrof\'isica de Andalucia (IAA) under a joint agreement with the University of Copenhagen and NOT. This research is based on observations made with the NASA/ESA Hubble Space Telescope obtained from the Space Telescope Science Institute, which is operated by the Association of Universities for Research in Astronomy, Inc., under NASA contract NAS 5–26555. These observations are associated with program 16642.

\section*{Declarations}

\section*{Data availability}
All spectra and photometric survey measurements are available through the respective data archives, which are publicly available, or upon request to the authors. The observed RVs are published in Supplementary Data 1.

%Some journals require declarations to be submitted in a standardised format. Please check the Instructions for Authors of the journal to which you are submitting to see if you need to complete this section. If yes, your manuscript must contain the following sections under the heading `Declarations':

\begin{itemize}
\item Competing interests: None
%\item Ethics approval and consent to participate
%\item Consent for publication
%\item Data availability 
\item Materials availability: Correspondence and requests for materials should be addressed to James Munday
\item Code availability: 
The fitting package WD-BASS that was used to determine atmospheric parameters and radial velocities is accessible at https://github.com/JamesMunday98/WD-BASS
\item Author contribution:
J.M. and R.P. carried out most of the modelling and analysis and wrote the majority of the paper. S.S. performed spectral fitting for the ultra-violet data. A.S.R. performed MESA modelling for accurate stellar compositions. I.P. and P-E.T. supervised the project. D.J. played an integral part in obtaining spectra of the target. G.N., M.M., S.T., A.B. and T.C. provided much insight and many discussions throughout all stages of the project. All authors contributed with comments and the writing of the manuscript.
\end{itemize}

%\noindent
%If any of the sections are not relevant to your manuscript, please include the heading and write `Not applicable' for that section. 

%\bigskip

%\bigskip\noindent
%Springer journals and proceedings: \url{https://www.springer.com/gp/editorial-policies}

%\bigskip\noindent
%Nature Portfolio journals: \url{https://www.nature.com/nature-research/editorial-policies}

%\bigskip\noindent
%\textit{Scientific Reports}: \url{https://www.nature.com/srep/journal-policies/editorial-policies}

%\bigskip\noindent
%BMC journals: %\url{https://www.biomedcentral.com/getpublished/editorial-policies}
%\end{flushleft}

\section*{Appendices: Radial velocity measurements}
\label{sec:appendixRVmeasurements}

\begin{table}
    \centering
    \caption{A table of all observed RVs (without relativistic correction) and the mid-exposure time-stamps. Wavelength calibration errors were propagated. Error bars are given as the standard deviation of 1000 bootstrapping iterations.}
    \begin{tabular}{r|r|r|r|r|r}
         HJD$-2450000$ & RV$_1$ & $\Delta$RV$_1$ & RV$_2$ & $\Delta$RV$_2$ & Instrument\\
        UTC & [km\,s$^{-1}$] & [km\,s$^{-1}$] & [km\,s$^{-1}$] & [km\,s$^{-1}$] & \\
\hline
8588.734699  &  $-$43.83  &  10.22  &  128.60  &  10.26  &  ISIS \\
8588.739403  &  $-$48.34  &  13.12  &  126.60  &  7.82  &  ISIS \\
8591.687642  &  $-$47.74  &  4.63  &  137.97  &  3.93  &  ISIS \\
8591.698183  &  $-$42.98  &  5.32  &  139.73  &  4.46  &  ISIS \\
8591.745818  &  $-$24.42  &  4.93  &  117.45  &  7.48  &  ISIS \\
8591.752887  &  $-$13.34  &  3.91  &  113.95  &  4.84  &  ISIS \\
8645.660736  &  $-$46.33  &  4.10  &  139.79  &  3.97  &  ISIS \\
8645.670120  &  $-$41.48  &  4.08  &  139.02  &  3.52  &  ISIS \\
8731.364001  &  149.73  &  9.28  &  $-$27.17  &  13.79  &  IDS \\
8732.411857  &  95.08  &  11.14  &  23.01  &  11.46  &  IDS \\
8732.432828  &  112.69  &  14.73  &  13.43  &  19.28  &  IDS \\
8732.460152  &  146.81  &  5.43  &  $-$24.69  &  8.03  &  IDS \\
8732.481127  &  137.36  &  9.86  &  $-$21.20  &  12.24  &  IDS \\
8732.506595  &  139.55  &  7.35  &  $-$18.77  &  3.99  &  IDS \\
8733.496923  &  $-$13.87  &  9.40  &  110.97  &  8.23  &  IDS \\
8733.517893  &  13.12  &  11.25  &  103.55  &  11.63  &  IDS \\
8734.382747  &  136.13  &  10.99  &  1.58  &  10.44  &  IDS \\
8751.390254  &  124.27  &  25.16  &  15.48  &  35.06  &  IDS \\
8751.400810  &  75.08  &  31.95  &  39.30  &  22.87  &  IDS \\
8751.411366  &  111.57  &  19.70  &  24.70  &  35.63  &  IDS \\
8751.421919  &  103.05  &  36.85  &  13.66  &  29.44  &  IDS \\
10385.714514  &  121.92  &  11.34  &  $-$48.04  &  5.86  &  FIES \\
10389.693347  &  95.12  &  3.91  &  $-$14.90  &  19.24  &  FIES \\
10389.711256  &  100.95  &  8.79  &  $-$19.22  &  14.96  &  FIES \\
10401.639249  &  137.53  &  4.08  &  $-$30.59  &  2.21  &  FIES \\
10401.675067  &  137.67  &  5.83  &  $-$44.11  &  3.54  &  FIES \\
10401.692976  &  149.54  &  12.12  &  $-$50.05  &  8.55  &  FIES \\
10402.602016  &  $-$40.56  &  10.65  &  122.89  &  4.08  &  FIES \\
10403.596407  &  23.18  &  18.31  &  56.32  &  4.86  &  FIES \\
10405.726945  &  119.48  &  6.67  &  $-$14.62  &  6.83  &  FIES \\
10412.657048  &  $-$25.61  &  7.60  &  146.64  &  4.69  &  FIES \\
10416.579027  &  123.96  &  6.33  &  $-$18.52  &  6.26  &  FIES \\
10426.543453  &  135.26  &  12.18  &  $-$39.47  &  8.71  &  IDS \\
10426.561527  &  155.21  &  6.64  &  $-$32.92  &  5.31  &  IDS \\
10426.579516  &  147.89  &  8.65  &  $-$51.49  &  5.29  &  IDS \\
10426.597579  &  150.34  &  8.63  &  $-$53.96  &  7.91  &  IDS \\
10426.615569  &  142.23  &  8.13  &  $-$36.32  &  5.93  &  IDS \\
10426.633728  &  141.73  &  10.30  &  $-$27.52  &  5.79  &  IDS \\
10426.651797  &  121.16  &  10.04  &  $-$22.66  &  9.27  &  IDS \\
10426.675840  &  104.72  &  6.37  &  3.20  &  4.94  &  IDS \\
10426.693911  &  99.42  &  18.86  &  10.39  &  9.92  &  IDS \\

    \end{tabular}
    \label{tab:appendixRVs1}
\end{table}

\begin{table}
    \centering
    \caption*{continued...}
    \begin{tabular}{r|r|r|r|r|r}
        \vspace{0.05cm}\\
         HJD$-2450000$ & RV$_1$ & $\Delta$RV$_1$ & RV$_2$ & $\Delta$RV$_2$ & Instrument\\
UTC & [km\,s$^{-1}$] & [km\,s$^{-1}$] & [km\,s$^{-1}$] & [km\,s$^{-1}$] & \\
\hline
10426.711986  &  60.03  &  9.14  &  41.66  &  9.22  &  IDS \\
10426.730069  &  55.33  &  4.57  &  54.96  &  3.95  &  IDS \\
10427.568303  &  22.90  &  13.36  &  98.97  &  12.29  &  IDS \\
10427.586293  &  38.01  &  10.43  &  82.12  &  10.65  &  IDS \\
10427.604333  &  52.56  &  13.10  &  63.98  &  7.18  &  IDS \\
10427.640501  &  78.96  &  10.17  &  37.00  &  5.90  &  IDS \\
10427.658595  &  80.16  &  9.54  &  2.07  &  10.28  &  IDS \\
10427.677213  &  111.05  &  6.78  &  $-$2.50  &  9.76  &  IDS \\
10427.695279  &  119.40  &  9.22  &  $-$15.23  &  5.24  &  IDS \\
10427.713363  &  129.54  &  14.44  &  $-$16.45  &  9.41  &  IDS \\
10435.613760  &  56.04  &  4.41  &  55.55  &  3.10  &  FIES \\
10451.692321  &  5.16  &  4.73  &  83.75  &  7.77  &  FIES \\
10451.710229  &  $-$3.64  &  4.86  &  117.82  &  3.49  &  FIES \\
10462.684526  &  126.24  &  5.78  &  $-$20.07  &  5.89  &  ALFOSC \\
10463.515200  &  57.13  &  6.48  &  60.23  &  3.75  &  ALFOSC \\
10463.522815  &  53.00  &  8.06  &  57.95  &  5.32  &  ALFOSC \\
10463.530392  &  42.85  &  7.76  &  82.97  &  8.49  &  ALFOSC \\
10463.613866  &  $-$30.69  &  6.79  &  150.23  &  15.31  &  ALFOSC \\
10471.660727  &  142.69  &  1.09  &  $-$44.14  &  0.80  &  UVES \\
10471.669856  &  145.29  &  0.81  &  $-$44.88  &  0.94  &  UVES \\
10471.678957  &  143.42  &  1.04  &  $-$40.03  &  0.71  &  UVES \\
10471.688045  &  141.58  &  1.02  &  $-$37.81  &  0.75  &  UVES \\
10471.697140  &  136.39  &  1.12  &  $-$32.99  &  0.98  &  UVES \\
10471.706232  &  135.55  &  0.93  &  $-$26.67  &  0.93  &  UVES \\
10471.715322  &  129.76  &  1.02  &  $-$24.49  &  0.93  &  UVES \\
10471.724414  &  122.58  &  1.17  &  $-$16.54  &  0.89  &  UVES \\
10471.733505  &  117.61  &  1.09  &  $-$14.89  &  0.85  &  UVES \\
10471.742600  &  106.43  &  1.07  &  $-$4.46  &  0.97  &  UVES \\
10471.751691  &  104.14  &  1.27  &  3.99  &  0.94  &  UVES \\
10471.760789  &  94.76  &  1.39  &  12.92  &  0.94  &  UVES \\
10471.769880  &  84.35  &  1.31  &  18.86  &  0.92  &  UVES \\
10471.778969  &  79.40  &  2.45  &  30.92  &  2.07  &  UVES \\
10471.788065  &  67.00  &  4.46  &  40.27  &  6.17  &  UVES \\
10471.797156  &  62.90  &  3.46  &  48.17  &  1.98  &  UVES \\
10471.806251  &  56.81  &  11.41  &  56.94  &  16.47  &  UVES \\
10471.815355  &  48.61  &  7.41  &  60.71  &  9.05  &  UVES \\
10471.824448  &  37.51  &  8.51  &  72.91  &  6.31  &  UVES \\
10471.833536  &  27.02  &  1.78  &  80.67  &  2.15  &  UVES \\
10525.451278  &  60.95  &  5.62  &  60.27  &  3.17  &  FIES \\
10525.469185  &  57.98  &  3.66  &  63.47  &  4.66  &  FIES \\
10545.435565  &  $-$33.32  &  6.71  &  131.51  &  4.64  &  FIES \\

    \end{tabular}
\end{table}

%\section{Section title of first appendix}\label{secA1}

%An appendix contains supplementary information that is not an essential part of the text itself but which may be helpful in providing a more comprehensive understanding of the research problem or it is information that is too cumbersome to be included in the body of the paper.




%\end{appendices}


%\bibliographystyle{sn-nature}
\bibliography{sn-article}% common bib file

\end{document}